# Monitoring Information Quality within Web Service Composition and Execution

**Thanh Thoa Pham Thi, Markus Helfert**

School of Computing, Dublin City University

**Abstract**  The composition of web services is a promising approach enabling flexible and loose integration of business applications. Numerous approaches related to web services composition have been developed usually following three main phases: the *service discovery* is based on the semantic description of advertised services, i.e. the functionality of the service, meanwhile the *service selection* is based on non-functional quality dimensions of service, and finally the *service composition* aims to support an underlying process. Most of those approaches explore techniques of static or dynamic design for an optimal service composition. One aspect so far is mostly neglected, focusing on the output produced of composite web services. In this paper, in contrast to many prominent approaches we introduce a data quality perspective on web services. Based on a data quality management approach, we propose a framework for analyzing data produced by the composite service execution. Utilising process information together with data in service logs, our approach allows identifying problems in service composition and execution. Analyzing the service execution history our approach helps to improve common approaches of service selection and composition.

**Keywords** service composition, service orchestration, data quality, business rules, and business process

## 1 Introduction

Motivated by the idea of "assembling application components into a network of services that can be loosely coupled to create flexible, dynamic business processes and agile applications that span organizations and computing platforms" [16], Service Oriented Computing (SOC) paradigm provides approaches that enable flexible business collaboration and enterprise application integration. Web services are the key technology in SOC, in which services are considered as "autonomous, platform-independent entities that can be described, published, discovered, and loosely couple in novel ways" [16]. A service oriented application includes a service provider and a service requester. A service discovery agency (e.g. UDDI) may act as intermediate between provider and requester and provides functionality to promote available services. The service provider defines a service description and publishes it (to the agency). After retrieving a suitable service, the service requester is able to invoke that service [19]. In this regard, service composition encompasses the process of searching and discovering relevant services, selecting suitable web services of best quality and finally composing these services to achieve an overall goal that usually in a business context aims to support an underlying business process.

On reviewing prominent approaches for service discovery, it appears it mainly involves functional attributes of services advertised in the service description. These include service type, operation name, input/output data format and semantics [10]. In order to select suitable services, a form of quality of service (QoS) evaluation is usually used as approach for service selection among many services of similar functionality. In literature, many approaches have been proposed to measure QoS with non functional quality criteria. In this context, QoS dimensions often refer to non-functional criteria which include *execution price*, *execution duration*, *reputation*, *reliability* and *availability* [23]. The functional quality of composite service is usually related to the connection of input data and output data between component services. For instance, [11] developed a method to measure the semantic links between component services, [12] proposed to constrain the input data and the output data of services for the composition. Otherwise, it is often assumed that the service functions execute according to the stated and published service description.

However, this is not ensure all output data during an execution of a composite service based on an underlying process are correct as with any execution and operation, this may not be the case. In contrast to other research, we consider this problem in data quality perspective and provide a framework that can help to detect some of the problems during the execution of the services.

We focus our investigation at the service deployment and execution phase. Motivated by the similarity to an information manufacturing system we view a composite service as a process that produces information products. In order to produces high quality information, the manufacturing process needs to be of high quality. This view follows a data quality and information quality perspective [24], which has been defined by widely accepted data quality dimensions such as *accuracy*, *completeness*, *consistency* and *timeliness* [25]. Similar to the manufacturing of products, processes that produce poor quality data are manifested in unsatisfied user requirements. As a consequence, in order to monitor the quality of service composition, we need to evaluate the compliance of service composition and execution with user requirements. Thus, we need to observe the data production process in form of composite services.

In order to introduce and describe this concept, in this paper, we first investigate data quality issues in service composition and execution, and then propose a framework for analyzing data produced by services. We relate our framework to common workflow and service composition approaches. In order to realise the proposed framework, we have developed a service log mechanism that captures all data updates in a workflow system. By analysing the service log information we are able to identify services and service instances that cause poor data quality, and thus this helps to improve the service selection and composition.

The remainder of the paper is organized as follows. Section 2 presents quality issues in web service composition and execution. Section 3 discusses some related work. Section 4 presents our framework for analyzing data in service composition and execution. Section 5 deals with a case study and the evaluation of our framework. Section 6 presents some concluding remarks and outlines some opportunities for future work.

## 2 Quality Issues in Web Service Composition and Execution

The process of web service composition and execution (WSCE) consists of defining a workflow that realises the required functionality followed by its deployment and execution on a runtime infrastructure [1]. The tasks in the workflow are individual and mapped to single services. The composite service is an implementation of a *Business Process* or a workflow that describes user requirements. The service orchestration defines the order of single service execution and the order of messages exchanged between services in accordance to the workflow. The execution of the composite service accesses to *Data*. Figure 1 describes quality issues in WSCE context.

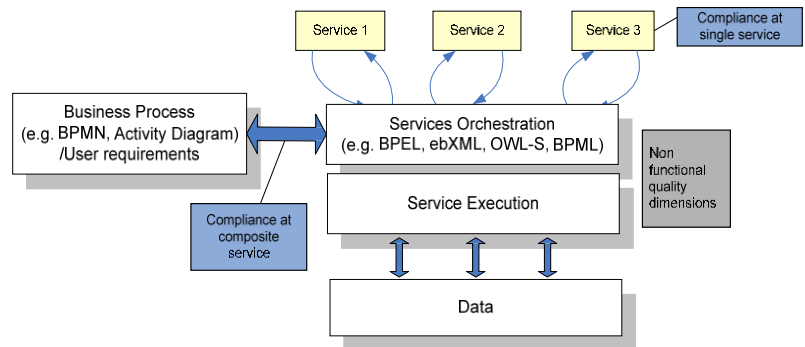

**Fig. 1** Functional and non-functional quality issues in WSCE context

It is widely recognised that the quality issues in web service composition and execution can be categorised in two dimensions: *functional* and *non-functional* quality. The functional dimension refers to the compliance of one individual service with its service description. This is extended with the compliance of a series of composite service with a defined workflow that captures the user requirements. The non-functional quality often addresses the performance of web services including execution price, execution duration, reliability, availability, and reputation [23]. A quality model for measuring quality of service composition with functional and non-functional levels has been developed in [11]. In which, the functional level mainly focuses on quality of semantic links between services. Two generic quality criteria considered for semantic links are *Common description rate* and *Matching quality* of the *Output* of the first service and the *Input* of the second one. Measuring the non-functional criteria focus on *Execution price* and *Response Time*.

Meanwhile, [6] proposes a conceptual model of Quality of E-Service with three dimensions *Environment quality*, *Delivery quality* and *Outcome quality*. Each dimension is refined with sub-dimensions, in which *Information Quality* is a sub-dimension of *Delivery Quality*. The authors consider *Information quality* as "the extent to which complete, accurate, and timely information is provided for the customer during the interaction process with the user interface".

Taking into account quality issues in WSCE, many approaches for static or dynamic web service composition and execution have been developed [1,14,17,22]. However, quality issues of mapping functional requirements to service composition and execution are current.

By adapting Data Quality Management perspective, a composite service can be considered as an information manufacturing process. Therefore, the quality of data produced by a composite service can be used to evaluate the functional quality of that service.

The data quality literature explored the most four dimensions of data quality which are accuracy, completeness, consistency, and timeliness [25]; Corresponding to the functional quality of WSCE, in this paper we explore three dimensions of data quality Accuracy, Completeness and Consistency (Table 1). Analyzing those dimensions help to clarify the causes of poor data quality in WSCE.

**Table 1.** Data quality issues in the context of web services composition and execution

| Quality dimension | Issues at Single service | Issues at Composite service |
|---|---|---|
| Accuracy, Completeness | - Incorrect input data, typo mistake;<br><br>- Incorrect output data due to inaccurate implementation of the functionality of the service | - Inaccurate orchestration of component services regarding the corresponding business process;<br><br>- Inaccurate implementation of composite service in transferring, calculating, or converting data received ( to transfer)  from (to) a service |
| Consistency | - Inconsistent data inside a single service | - Non uniform output data of a service to be directly inputted to the following service |

## 3 Some Web Service Composition Approaches Taking into Account Functional Quality

Today, many web service composition languages have emerged such as BPML[3], OWL-S and BPEL4WS [2]. The later, BPEL4WS is a popular language that defines service composition on the basis of a business process. It includes data flow controls among services to be composed. In addition, business rules are a typical used to constrain the service composition and ensure integrity checks during the execution.

Regarding functional validation of composite services, [12] proposes a proactive approach to ensure data quality between composite service by using data constraints. Data constraints enforce the correctness of data transformation between services. This approach also develops a method to regulate data constraint outside the service needed for the composition to fill semantic gaps in data transformation that may occur in service reusability and composability.

In [21], the authors have proposed Petri-Nets based approach to validate the connectivity between single services for a service composition. Meanwhile, [5] deals with developing a hybrid composition approach to control data flow between single services in accordance to the underlying process. The approach separates business policies and constraints from the core composition logic. This allows to increase the flexibility of composition. At a single service level, Functional Quality of Service (FQoS) metrics have been proposed measuring the similarity-based approximate matching between a query and service description [10]. In addition, [20] proposes a rule-based approach to handle service composition life-cycles to support dynamic binding and flexible service composition. Each phase in the life cycle has related rules that need to be taken into account and help to ensure the integrity. The approach takes into account business rules as well as constraints on resource, time and costs.

These approaches are mainly based on business rules and constraints designed at the service composition phase. However, these approaches are limited to track and monitor service composition and execution often resulting in data quality problems. Furthermore they do not investigate the correctness and effectiveness of the service composition after the services are executed. As a consequence, most approaches are unable to detect divergence of services execution form user requirements.

## 4 Framework for Analyzing Data Quality

The proposed framework follows a business rules-based approach to data quality. Business rules are "statements that defines or constrains some aspects of a business" [18]. Aspects include the business policy, the business structure, the organization structure, business objects, and business process. A business rule is normally described by the natural language, which can be reconstructed in form of Event-Condition-Action [4, 8], in form of If-Then expression, or in form of <Subject> Must <constraints> [13]. In addition, business rules can be expressed in formal languages such as the logic predicate, or the first logic order language or pseudo SQL statement to facilitate business rule implementation. In data quality, business rules related to business objects or data objects as well as business processes.

Base on common business rule approaches, we developed a framework for WSCE that is presented in Figure 2. The main advantages of this framework are as follows:

- Monitoring the mapping of selected services with tasks of a specific conceptual business process.

- Monitoring the service execution and data produced
- Consequently, we are able to identify services causing data quality problems.

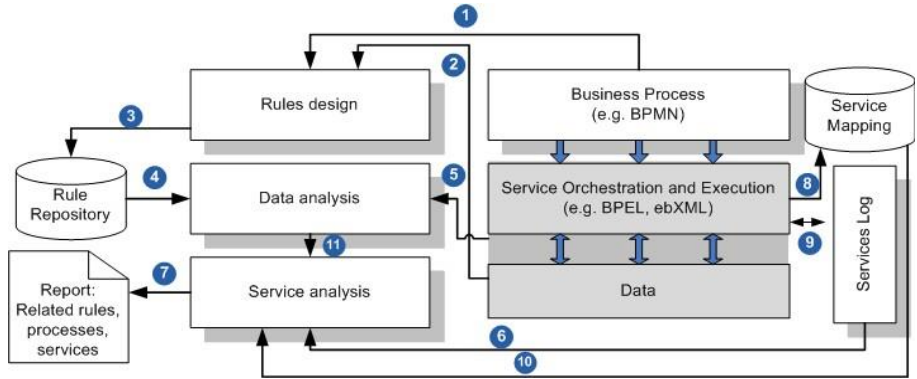

**Fig. 2**  Framework for monitoring data quality within WSCE

Figure 2 depicts the proposed framework. The *Service Orchestration and Execution* component and *Data* component are out of our framework. The detail of other components and relationships between them described by flows of data are explained in the following.

*Business Process:*

This block represents the conceptual business process (BP) model that a service composition must comply with. A BP model can be described with BPMN, UML or EPC model. In our framework, we are interested in the input and output of data of activities/tasks in the BP. A part of the meta-data of the BP model is stored in the Service Mapping and Rule Repository (Figure 3). The business process model and Data components are the basis for business rules design (Flow 1 and Flow 2).

*Business rules specification*:

This component concerns specifying business rules. A business rule is related to one or many activities/tasks in a business process and/or data objects. Business rules can be designed manually based on business expert's experiences. However, some complex rules such as dynamic rules, which relate to status change of data objects, can be detected and designed with help of some techniques [26].

Rule specification language is a pseudo logic predicate language which can be automatically converted into code for analyzing data. The rule is usually in form of If *<Boolean logic expression>\** Then *<Boolean logic expression>\**, or an assertion of *<Boolean logic expression>\**.

*<Boolean logic expression>** is composed of one or many <Boolean logic expressions> combined together with logical operators. A *<Boolean logic expression>* = *<left expression >* *<comparison operator>* *<Right expression>*.

*<left/right expression>* is a mathematical expression including data attributes, data values, mathematical operations, or aggregation functions.

The specified rules are stored in a *Rule repository* (Flow 3) such as relational databases or XML files. The *<Boolean logic expression>** is described with the binary tree structure (Figure 3). The *Expression node* class presents a data attribute, or data value, or an operator, or a function. The *NodeRelation* class expresses the relation between a parent node and a child node within left/right position.

*Service Mapping repository*:

This repository captures the mapping between services to be composed to tasks/activities defined in the underlying business process. A task corresponds usually to a service, and a service can correspond to one or many tasks. However in the case that a task corresponds to many services, the service composition is significantly more complex and is currently not subject of this study. The information is usually stored within the service composition phase (Flow 8). Figure 3 depicts the structure of the *rule repository* and service *mapping repository*. The links between Service, Activities and Process describes mapping information.

*Service Log:*

The service log captures specific events occurred during the service execution. In our framework we are particularly interested in events related to data updates and changes. It is necessary to record what service instance in what composite service instance writes what data to the database. A service needs to have information where to write the data, in other words, information about database server, table, and column are known. Furthermore, the record ID is also captured after writing the data. All these information should be stored in the service log (Flow 9).

Although there are approaches to specify log formats [7], current approaches such as the Common Log Format [9] and Combined Log Format of W3C are not sufficient for our approach as they are not directly able to represent the required information. Therefore, we propose a practical oriented log file. The log file entries contain following information:

```
<Entry>
  <Service location> …</Service location> // registry location
  <Service ID>… </Service ID>
<Service instance ID>… </Service instance ID>
<DbServerID>… </DbServerID>
<TableName>… </TableName>
<ColumnName>… </ColumnName>
<RecordID>...</RecordID>
<Data Value>...</Data Value>
</Entry>
```

The *RecordIDValue* contains the value of the primary key attributes of a record in the table *Table Name* in the data base *DbServerID*. Data Value is the value written to the DB at the column *Column Name*. The primary key attributes of a table is identified based on the meta-data of the data base.

*Data analyzing:*

This component analyzes data in the database against rules defined in the rule repository (Flow 4 and Flow 5). Data that violates some rules will be marked. Information on its Table, Column, and RecordID is reported. Furthermore, based on the information in the rule repository (Figure 3), the task(s) related to these rules are also identified and reported. Each line in the report is composed of the following information: DbServerID, TableName, ColumnName, DataValue, RecordID, RuleID, and TaskID.

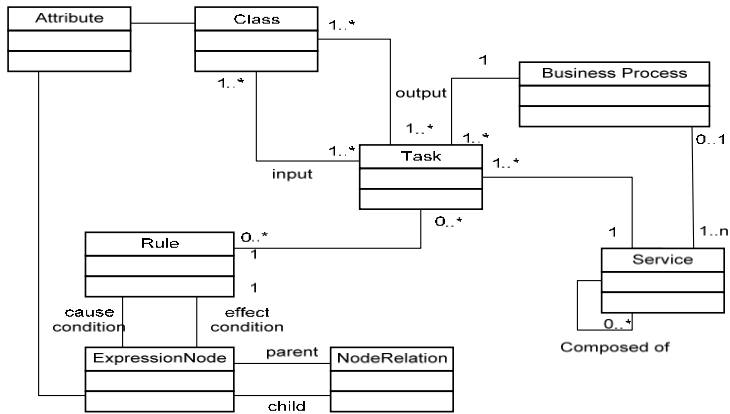

**Fig. 3** Rule and service-process mapping data model

*Service analyzing*:

This component aims to identify services related to poor data quality. From the data analysing result (Flow 11) and the mapping information between Task and Service in addition to the service logs (Flow 6 and Flow 10), we are able to identify particular service instances that caused poor data quality.

In order to deploy our framework we divided the framework into two phases: *preparation* and *analyzing*. The preparation phase concerns with implementing service logs, and recording information of the mapping between services and business processes. The analyzing phase addresses the modelling of business rules and then subsequently analyzing data against these rules.

# 5 Case study and Discussion

The following case study illustrates our framework with a common service that provides the booking of travelling packages. Initially, customers search information about available travel packages, and then subsequently may book a flight and a hotel. Once the booking is completed, the user pays the total amount and confirms the booking. Alternatively the user may cancel the booking. We developed a service oriented application for Travel package booking and analyzing data quality of the application along the two phases: preparation and analyzing.

In the *preparation* phase, initially following *Business Process* block in the framework, a conceptual business process model for the booking travelling package is modelled (Figure 4). Next, the service composition process is realized (*Service Orchestration and Execution* block); the mapping information between the tasks of the BP and individual services are stored (*Service Mapping* repository). We suppose this is a design time service composition. The service composition can be described with BPEL based on the orchestration depicted in Figure 5.

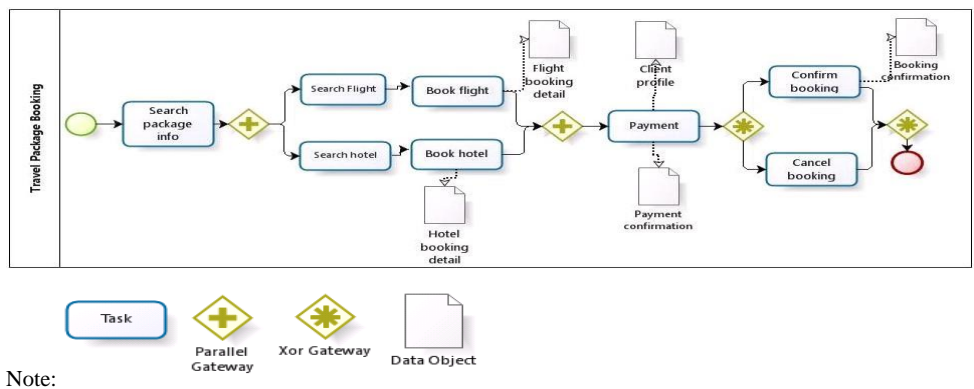

Note:

**Fig. 4** Business process of Travel package booking described with BPMN

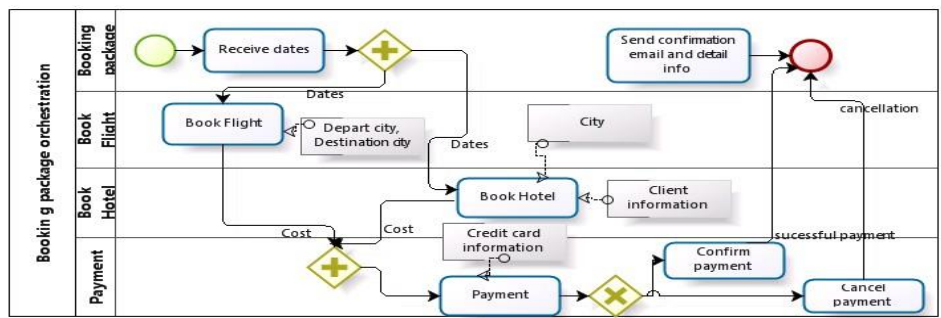

**Fig. 5** Service orchestration

The composite service *Booking package* is composed of a set of available services: *Book Flight* service, *Book Hotel* service, and *Payment* service. The flows of data/message between services are also described in the orchestration.

Mapping services and tasks in BP are as following table:

| Task | Service |
|------|---------|
| Search Flight | Book Flight |
| Book Flight | |
| Search Hotel | Book Hotel |
| Book Hotel | |
| Payment | Payment |

Other tasks are realised at the composite service levels such as *Search package information*, *Send confirmation email*, etc. The service log techniques are implemented to record data updates events (*Service log* repository). The database schema of the application is described in Figure 6.

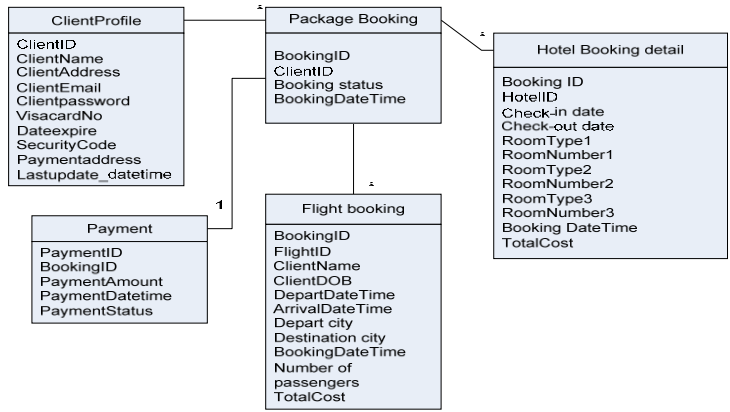

**Fig. 6** Database schema of Travel Booking

Once our framework was set up and developed in the preparation phase, subsequently we use our framework in the *analyzing* phase. After deployment of the framework, the data in database and service logs serve for the analyzing. First, business rules are designed in regard to the business process model and data schema depicted in Figure 4 and Figure 6.

For our case scenario, we suppose there are the following rules designed:

**R1:** If the booking is confirmed then the payment must be full paid. This rule relates to *Payment* task and *Confirm booking* task and is described with pseudo SQL as follows:

```
If the PackageBooking.BookingStatus = 'confirmed' then
Payment.PaymentStatus='Full Paid' AND
PackageBooking.BookingID=Payment.BookingID
```

**R2:** For every booking, the payment fee must be equal to the total cost of Flights, Hotels and including all tax. This rule relates to *Payment* task, *Book Flight* task, and *Book Hotel* task and is described in pseudo SQL:

```
Payment.PaymentAmount=HotelBooking.TotalCost +
SUM(Flightbooking.FlightCost) AND
Payment.BookingID=Flightbooking.BookingID AND
HotelBooking.BookingID=Payment.BookingID
```

Once the rules are set up are stored in *Rules repository*, the data in the database is analyzed against these rules (*Data analysis* blog). These rules can be automatically generated into code to verify data in database. The data analyzing component is able to distinguish main conditional expressions and join expressions in a rule.

A join expression links foreign keys and primary keys attributes of two tables. For example, for rule R1, the condition is:

```
If the PackageBooking.BookingStatus = 'confirmed' then
Payment.PaymentStatus='Paid', the join expression is
PackageBooking.BookingID=Payment.BookingID.
```

The data analyzing focuses on the condition expressions; possible attributes related to data errors regarding the two rules are

```
PackageBooking.BookingStatus, Payment.PaymentStatus,
Payment.PaymentAmount, HotelBooking.TotalCost, and
Flightbooking.FlightCost
```

If there is any data that violates a rule, then the related process will be identified based on information stored in the rule repository. For example, suppose there are incorrect data listed in Table 2. Also, assume there is a confirmed Booking however which is not fully paid. This relates to two records; one in the Booking table and the other in the Payment table. The rule R1 is related to the tasks Payment and Confirm booking.

**Table 2.** Analyzed data report

| DBServerID | Table Name | Column Name | Data Value | RecordID | RuleID | TaskID |
|---|---|---|---|---|---|---|
| DB1 | Booking | BookingStatus | 'confirmed' | 1200 | R1 | Payment, Confirm booking |
| DB1 | Payment | PaymentStatus | 'unpaid' | 3200 | R1 | Payment, Confirm booking |

Next, the *Service analysis* is realised. The service log file describes the data of the `BookingStatus column, RecordID 1200` is produced/written by the composite service *BookingPackage*, meanwhile the data of `PaymentStatus column, RecordID 3200` is produced/written by the service *Payment*.

**Table 3.** Extracted service log information

| Service location | Service ID | Service instance ID | DB ServerID | Table Name | Column Name | Record ID | Value |
|---|---|---|---|---|---|---|---|
| url:// | BookingPackage | 10 | DB1 | Booking | BookingStatus | 1200 | Confirmed |
| url:// | Payment | 12 | DB1 | Payment | PaymentStatus | 3200 | Unpaid |
| url:// | BookFlight | 5 | DB1 | | | | |

From the above information, we can identify that the cause of the incorrect data related to Table 2 is the Payment service and the BookingPackage composite service.

The case study is motivated on a real world application, however significantly simplified and abstracted. Also, we have assumed that key techniques have been implemented within the realistic scenario in order to apply our approach. For example techniques to store the mapping of service and tasks of a conceptual business process model at run time service composition, as well as techniques to record and manage the service log. With the case study we illustrated how our framework can be used. Discussion with practitioners, our approach has advantages for identifying problems with service execution and thus resulting in unsatisfied user requirements (i.e. caused by poor data quality), and it is also needed to conduct a performance evaluation of our approach in the future.

# 6 Conclusion

Motivated from a data quality perspective on service composition, in this paper we have presented a framework for monitoring web service composition and execution; in particular, we investigate if the service and composite service execution comply with user requirements. User requirements are represented in form of business process and business rules. Our framework follows a data quality management approach and incorporates business rules concepts. We propose that the service composition is based on business processes, and the mapping between tasks in business process and services are recorded. We also propose a new service log format to support the analyzing. Our approach differs from others prominent approaches by investigating service deployment and data produced by services. We have developed a tool for data quality analyzing and we illustrated the framework within a case scenario that shows that our approach can assist to improve service selection and composition. We have developed business rules specification and data analyzing.

In the future we aim to extend and develop techniques and tools to support and improve the proposed framework such as techniques to handle service logs, and analyzing services, as well as evaluating the performance of the proposed approach.